# A linear theory underlying quantum mechanics.

Casey Blood
Professor emeritus, Rutgers University
CaseyBlood@gmail.com

## Abstract

Linearity allows several versions of reality to simultaneously exist in the state vector. But it implies that there is no interaction between versions, and that there will never be perception of more than one version. It also implies, in conjunction with group representation theory, that the particle-like properties of mass, energy, momentum, spin and charge are attributes of the state vectors. These results can be used to show there is no evidence for the objective existence of particles. The properties of the wave function are sufficient to explain all the particle-like properties of matter.

Representation theory is also extensively employed in the Standard Model, with gauge fields transforming as representations of the internal symmetry group. And when applied to the permutation group, it is essential for understanding symmetric and antisymmetric states. In fact all of quantum mechanics is set up exactly as if it were the representation of an underlying pre-representational theory. A linear equation structure for the underlying theory is suggested, and it is shown in outline how quantum field theory emerges as a representational form of the pre-representational theory.

## 1. Introduction.

Quantum mechanics is an extremely successful theory, with its mathematics giving a correct description of phenomena from elementary particles to atoms to semiconductors to the structure of stars. But it has a most peculiar characteristic. Because of its linearity, the state vector can contain more than one version of reality—Schrödinger's cat is simultaneously dead and alive for example. It is often suggested that this implies pure linear quantum mechanics, by itself, cannot account for our perception of only one version. That, however, is not correct; linearity implies only one version of reality is perceived. In addition, linearity implies that if a wave function is spread out over many grains of film, only one grain will be perceived as exposed. More generally, linear quantum mechanics, as we show in Secs. 2 through 6, implies that all the perceived particle-*like* properties of matter can be accounted for solely by the properties of the state vectors, with no assumption of the separate existence of particles. That is, wave-particle duality is simply a duality in the properties—wave-*like* and particle-*like*—of the wave function [1].

Another remarkable characteristic of quantum mechanics is that virtually the whole theory is couched in the language of group representation theory. Particle-like states, represented by kets $|m,E,p,S,s_z,Q\rangle$, are labeled entirely by the group representational quantities of mass, energy, momentum, spin and charge. In addition, symmetric and antisymmetric states correspond to one-dimensional representations of the permutation group. And properties of representations of the internal symmetry group are employed in a number of places in the Standard Model.

A natural way to obtain this ubiquitous group representational structure is to suppose the current theory is a representational form of an underlying, pre-representational linear equation. To illustrate this idea, we suppose the underlying theory consists of a single, linear partial differential equation in some currently unknown set $U_i$ of underlying variables [2,3]. To match the known group structure, it is assumed the linear operator, $O(U_i)$, is invariant under a group of transformations of the $U_i$ which is homomorphic to the direct product of the Lorentz group, an internal symmetry group, and a permutation group. Neither space nor time nor the particle-like properties of mass, spin and charge enter the equation; these all correspond to properties of its solutions. All physical states—the vacuum, fermions, and bosons—are presumed to correspond to antisymmetrized solutions of this single equation. A relatively simple example is given in the first part of Sec. 7. The general problem is then considered, with a sketch of how one goes from the pre-representational form to the current, representational form of quantum field theory by using a set of spin ½ particle-like basis vectors.

## 2. Perception of only one outcome.

The linearity of the quantum mechanical equations allows several versions of reality to simultaneously exist. But as we will show using the half-silvered mirror experiment as an example, it implies that more than one version is never perceived. Light from a source is shone on a half-silvered mirror. Half the beam is transmitted through the mirror in a horizontal beam and half is reflected in a vertical beam. Detector H is on the horizontal path and detector V is on the vertical path, with the detectors reading either no (no detection) or yes (detection). There is also an observer who observes the states of the detectors and writes what she perceives. We shoot a single photon (photon-like wave function) at the mirror. At time 0, after the mirror but before detection, the state vector is

$$|\Psi, 0\rangle = a(H)|\text{version 1 of reality, time 0}\rangle + a(V)|\text{version 2 of reality, time 0}\rangle$$

(1)

$|\text{ver. 1,0}\rangle = |\text{photon wf on H path}\rangle|\text{Det H,no}\rangle|\text{Det V,no}\rangle|\text{Obs writes "I see no,no"}\rangle$
$|\text{ver. 2,0}\rangle = |\text{photon wf on V path}\rangle|\text{Det H,no}\rangle|\text{Det V,no}\rangle|\text{Obs writes "I see no,no"}\rangle$
$|a(H)|^2 + |a(V)|^2 = 1$

where a(H) and a(V) are coefficients. Note that *in the mathematics of quantum mechanics*, there is no "the" photon, and as we will see in a moment, no "the" detector, and no unique "you;" instead there are multiple **versions** of each "object."

A linear time translation operator U(t) takes the system from time 0 to time t;

$$|\Psi, t\rangle = U(t)[|\Psi, 0\rangle]$$
$$= a(H)U(t)[|\text{ver. 1,0}\rangle] + \qquad (2)$$
$$a(V)U(t)[|\text{ver. 2,0}\rangle]$$

Because U(t) acts only on [|ver. 1,0⟩] in the second line—that is, a(H) and a(V) are not inside the square brackets—its effect on |ver. 1⟩ must be independent of a(H) and a(V). Thus U(t)|ver. 1,0⟩=|ver. 1,t⟩ can be calculated at a(H)=1, a(V)=0. That is, independent of a(H) and a(V), the horizontal version evolves in time as if the vertical version were not there (because one

can assume in calculating it that a(V)=0). This implies no photons from the versions of the detectors on the second branch can ever reach the version of the observer on the first branch. This is the principle of the **isolation of branches**. It implies the versions of the observer can perceive only what occurs on their own branch of the state vector. And it holds even if the time translation operator is not unitary.

We therefore have, with the photon being annihilated,

$|\Psi, t\rangle = a(H)|\text{version 1 of reality, time t}\rangle +$
$\qquad a(V)|\text{version 2 of reality, time t}\rangle$

(3)

|ver. 1,t⟩=|Det H,yes⟩|Det V,no⟩|Obs writes "I see **only** yes, no" ⟩
|ver. 2,t⟩=|Det H,no⟩|Det V,yes⟩|Obs writes "I see **only** no, yes" ⟩

Because each version of the observer can perceive only what occurs on her own branch, each version will write that she saw only one version of reality. No version writes "I see something other than a single, 'classical' version of reality."

It is critical to note that Eq. (3) follows from the allowed interactions. U(t) allows only the H version of the photon to change the H detector from no to yes. And it allows only the "messenger" photons from the |Det H,yes⟩|Det V,no⟩ version of the detectors to activate the eye-brain- hand written response "I see **only** yes, no." (And similarly for the V version.) Thus **the dynamics** of the detectors, the light that goes from the detectors to the eye, the observer's eye, brain and hand **absolutely dictate this written result**. No basis—in particular no "preferred" basis—was used in writing Eq. (3), so no preferred basis was used in showing that only single, classical versions of reality are perceived in quantum mechanics.

## Generalization.

This can be generalized. Linearity dictates that the versions of the observer never receive sensory input from any version of reality other than their own. Thus quantum mechanics implies that, even though more than one version of reality may "exist" in the wave function, no version of the observer ever writes "I see something other than a single, classical version of reality." Thus, because that is never written in quantum mechanics, **more than one version of reality is never communicably perceived** in quantum mechanics, no matter what basis you choose. And if more than one version is never *communicably* perceived, then for all practical purposes, more than one version is never perceived. That is, non-communicable sensory perceptions, whatever they might be, are not relevant here.

## Two observers never disagree.

If we add a second observer to the versions 1 and 2 of Eq. (3) and instruct the second observer to write whether she agrees with what the first observer wrote, the result will be

|ver. 1,t⟩=|Det H,yes⟩|Det V,no⟩|Obs 1 writes "I see **only** yes, no" ⟩
$\qquad$|Obs 2 writes "I see yes, no, in agreement with Obs. 1." ⟩ (4)
|ver. 2,t⟩=|Det H,no⟩|Det V,yes⟩|Obs writes "I see **only** no, yes" ⟩
$\qquad$|Obs 2 writes "I see no, yes, in agreement with Obs. 1." ⟩

Thus two observers never communicably disagree because the versions on each branch can perceive and communicate only what is on that branch.

## 3. Parts of a wave function.

In classical physics, a small part of a wave, a water wave for example, carries only a correspondingly small part of the energy of the wave. But this statement is misleading in quantum mechanics. To see this, we continue to use the half-silvered mirror experiment. Then the linear energy operator $\mathcal{E}$, applied to the wave function after the mirror, gives

$$\mathcal{E}[|ph\rangle]=$$
$$\mathcal{E}[a(H)|ph\ H\rangle+a(V)|ph\ V\rangle] \quad (5)$$
$$a(H)\ \mathcal{E}\ [|ph\ H\rangle]+a(V)\ \mathcal{E}\ [|ph\ V\rangle]$$

with "ph" standing for photon. If we now assume |ph> is an eigenstate of the energy with eigenvalue E both before and after the mirror, then

$$E[a(H)|ph,\ H\rangle+a(V)|ph,\ V\rangle]=a(H)\ \mathcal{E}\ |ph\ H\rangle+a(V)\ \mathcal{E}\ |ph\ V\rangle \quad (6)$$

We set $a(H)=\exp(i\theta)|a(H)|$, $a(V)=\exp(i\phi)|a(V)|$ and note that $|ph, H\rangle$, $\mathcal{E}\ |ph, H\rangle$, $|ph, V\rangle$, $\mathcal{E}\ |ph, V\rangle$ and E are all independent of $\theta$ and $\phi$. Then taking the derivative of Eq. (6) first with respect to $\theta$ and then with respect to $\phi$ shows that

$$\mathcal{E}|ph,\ H\rangle=E|ph,\ H\rangle \quad (7)$$
$$\mathcal{E}|ph,\ V\rangle=E|ph,\ V\rangle$$

That is, the two kets after the mirror are also separately eigenstates of the energy, with the same eigenvalue as the original state. (Note, however, that if the experiment is repeated many times, the horizontal branch will carry an *average* energy of $|a(H)|^2 E$ because of the probability law.) This property—that each branch carries the full energy (and momentum)—is necessary for the next section, where a small part of the wave function triggers a film grain, and also for the Compton and photoelectric effects.

## 4. Localized perception from a spread-out wave function.

In the half-silvered mirror experiment of Sec. 2, the results exactly imitate what we would expect if there really were an objectively existing, particulate photon. The versions see either |Det H,yes⟩|Det V,no⟩, as if the alleged photon had traveled on the H path, or |Det H,no⟩|Det V,yes⟩, as if the photon had traveled on the V path; and no version ever perceives |Det H,yes⟩|Det V,yes⟩. What we will show here is that this imitation of the particle-like property of localization—that there is perceived detection in one and only one, localized place—holds in all cases.

We use a scattering experiment to illustrate. A single electron (electron-like wave function) scatters off a proton (proton-like wave function) and hits a screen coated with N grains of film. Experimentally we know that if the grains are analyzed, only one will be found to be exposed, even though the wave function hits them all. The goal is to show that quantum mechanics implies only one grain will be *perceived* as exposed. Neither the existence of particles nor collapse (nor an argument involving decoherence or quantum Darwinism [4-7]) is necessary to obtain this perceptual result.

After the slit but before detection, the state vector of the electron is

$$\int d^3x\, \psi(\mathrm{x})|\mathrm{x}\rangle \tag{8}$$

where $\psi$ is the wave function. And the state vector of the N grains is

$$\prod_{j=1}^{N} |gr\, j\rangle \tag{9}$$

As the electron wave function passes through the layer of grains, one can divide the integral of Eq. (8) into a sum, with one term for each grain. We suppose grain j has frontal area $\Delta A_j$ and take $z_j$ as the direction perpendicular to $\Delta A_j$. Then

$$\int d^3x\, \psi(\mathrm{x})|\mathrm{x}\rangle = \sum_j \int dz_j \int dA_j\, \psi(x)|x\rangle \tag{10}$$

For each grain $j$, there is an interaction term $H(j)$ such that if the electron hits that grain, it gets changed from non-exposed to exposed. Thus the $j^{th}$ term in Eq. (10) will expose grain $j$ and only grain $j$. This implies that after the electron wave function passes through the crystals, the full wave function, electron plus grains, will be a sum,

$$|\Psi\rangle = \sum_{j=1}^{N} |gr\, 1\rangle \ldots |gr\, j^*\rangle \int dz_j \Delta A_j \psi(x)|x\rangle \ldots |gr\, N\rangle \tag{11}$$

with one and only one exposed grain, indicated by the asterisk, in each term.

Each of these N terms corresponds to a separate version of reality. If we now include an observer who looks at the state of each grain, there will be a different version of the observer for each of the N versions of reality. And in accord with our principle of the isolation of branches, the version on branch $j$ can perceive only the state of the grains on version $j$. Thus when the observer is included, the state vector becomes

$$\sum_{j=1}^{N} |gr\, 1\rangle \ldots |gr\, j^*\rangle \int dz_j \Delta A_j \psi(x)|x\rangle \ldots |gr\, N\rangle |\text{Obs sees only grain } j \text{ exposed}\rangle \tag{12}$$

So we see that, in spite of the fact that the electron wave function is spread out over all the grains, there will never be **perception** of more than one exposed grain. (Note: The small amount of the wave function that hits each grain carries the full energy of the electron, in accordance with Sec. 3, so it has sufficient energy to expose the grain.)

The result is that it is not necessary to assume either collapse or the existence of particles to explain why only one **localized** grain is perceived as exposed. The mathematics of quantum mechanics, by itself is sufficient to explain the perception of a localized effect from a spread-out wave function.

### Trajectories.

An extension of this argument can be used to show that linear quantum mechanics predicts the perception of sharply defined, particle-like trajectories. Put several layers of film grains directly behind the first and assume the electron goes straight through the grains. Then because each branch of the electron wave function becomes localized on the trajectory defined by the initial grain it exposes in the first layer, there are still only N outcomes, and in each outcome, only those grains directly in line are exposed. Thus one ends up with N versions of reality, each having a straight-line trajectory of exposed grains. Allowing the electron to scatter to the side a little increases the number of possible trajectories, and they may be curved, but each trajectory will still be perceived as an essentially continuous, sharply defined line of exposed grains. There will never be a version of the wave function in which there is a disjoint or diffuse trajectory and so no version of the observer will perceive anything other than a classical, particle-*like* trajectory.

## 5. The particle-like properties of matter: mass, energy, momentum, spin and charge.

The five particle-*like* properties of mass, energy, momentum, spin, and charge have been associated with objectively existing particles since the time of Newton. But they are actually properties of the state vectors because they can be **derived** from linearity and the invariance principles of quantum mechanics. This derivation of the properties that, along with localization, essentially define particles is a strong (but not quite rigorously unassailable) argument that linearity is an absolute, basic principle of the mathematical theory of matter.

### Classifying solutions to equations in quantum mechanics.

The mathematical reasoning starts with a property of solutions to calculus equations. In algebraic equations, such as $3x+4=10$, there is often only a single solution. But almost all the equations in quantum mechanics are calculus equations, and they typically have a continuous infinity of solutions. This means that, to avoid chaos, it is often important to find some way of classifying solutions to calculus equations.

The classification process for quantum mechanics is most interesting.
(1). It starts from the **physical observation** that the outcome of experiments on the atomic level does not depend on how the experimental apparatus is oriented. It can be oriented North-South, East-West, up-down, or at some angle between these, and that makes no difference to the outcome of the experiments (provided gravity can be ignored).
(2). This physical observation has a reflection in the **form of the equations**, say the Schrödinger equation for the hydrogen atom. They must have a certain symmetry (They "look the same" when one changes the orientation of the axes.)
(3). The form dictated by the symmetry then has a consequence for the **classification** of solutions. In the case where the orientation of the apparatus in three-dimensional space makes no difference, the form of the equations implies solutions can be classified according to their angular momentum. This classification process uses group representation theory which is outlined in Appendix B.

(4). The classification is not just mathematical; it has **physically measurable consequences**. The electron-like wave function in the hydrogen atom has measureable angular momentum.

(5). Finally, the mathematics yields a **conservation law**; in the hydrogen atom example, the actual numerical value of the total angular momentum stays the same no matter what happens within the system.

There is an interesting consequence of the classification scheme in the case of angular momentum. The mathematics tells us, and experiment confirms, that in the hydrogen atom case it can have only integer values, 0, 1, 2, … (in appropriate units). That is—in contrast to the classical case, where angular momentum can take on any value—angular momentum (spin) is *quantized* ; it can take on only certain discrete values rather than being able to take on a continuum of values!

It is amazing that one can get a physically measurable property and a conservation law out of almost nothing—just linearity and the fact that the orientation of the apparatus makes no difference! And not only that, but one can (correctly) deduce the *quantization* of angular momentum; it can only take on the values 0, 1, 2,… This is a great example of Wigner's "unreasonable effectiveness of mathematics."[8]

### Classification of particle states from space-time symmetries.

We can generalize from the hydrogen atom example in which only invariance to rotation in three dimensions was taken into account. First, experimental results do not depend on *where* the experiment is done, or *when*. Further, Einstein deduced relativistic invariance—that the equations of physics must be invariant under certain "rotations" involving both space and time (these include rotations in three-dimensional space, but they also involve velocity). When combined with linearity, these invariances—where, when, four-dimensional relativistic rotations—then have a number of consequences for classifying the solutions to the equations.

First, the invariance under *where* and *when* implies the solutions—the state vectors— can be labeled by **energy and momentum**. Further these **mathematical labels** correspond exactly to the **physically measurable properties** of energy and momentum. In addition, one gets **conservation** of energy and momentum; for an isolated system, the total energy and momentum remain the same forever.

Second, if we also take into account invariance under Einstein's relativistic rotations, then the solutions acquire two additional properties—**spin** (analogous to the angular momentum of the hydrogen atom) and **mass**. In this case, the allowed values for spin are still quantized, but they also include *half*-integer values; 0, 1/2, 1, 3/2, 2,… (The possible values for mass are not limited by the invariances).

This is even more amazing! Mass itself, perhaps the most basic property of matter, is a consequence of the simple observation that the outcomes of experiments don't depend on the orientation (generalized to include relativistic rotations) or position of the apparatus! And the mathematically allowed quantized values for the angular momentum—and *only* these values— are exactly observed; electrons, quarks and neutrinos, for example, have spin ½ while the photon has spin 1 (and no particle has, for example, spin ¾).

## Internal symmetries and charge.

The last basic property of matter is charge. The charges have the same invariance-under-rotation type of origin as spin, only in a somewhat peculiar way. It was found experimentally that the proton, neutron and all other strongly interacting "elementary" particles were made up of three more elementary particles—quarks. The only way to make sense of all the various experiments was to suppose the quarks came in three "colors" (nothing to do with actual colors, of course; just colorful language). Further experiment showed that the theory had to be invariant—did not change form— under rotations in the color space, just as the hydrogen atom theory was invariant under rotations in our ordinary three-dimensional space. And the group theoretic *labels* associated with this invariance are the strong charges [9].

A similar argument can also be made for the electromagnetic and weak charges, with all charges quantized as integer multiples of the basic charges. (In the case of quarks, the basic electrical charge is 1/3 the charge on the electron.) Thus, just as mass and so on are labels, and physical properties, associated with the wave function due to invariance under space-time transformations, so the three types of charges are labels, and physical properties, associated with the wave function due to an internal—"within the particle," so to speak; not in extended space and time—set of rotations in some abstract space.

## Addition.

We also note that if there are several "particles" present, group representation theory correctly predicts that the energies, momenta, z components of spin, and charges all add algebraically. And it correctly predicts the somewhat more complicated addition rules for total spin (angular momentum).

## Symmetry and antisymmetry.

Group representation theory is also related to the antisymmetry of fermion states and the symmetric statistics of bosons states. Suppose we have a linear equation

$$O\bigl(u(1),\dots,u(n)\bigr)\psi\bigl(u(1),\dots,u(n)\bigr) = 0$$

where $O$ is a linear operator, and suppose that $O$ is invariant under all exchanges $u(i) \leftrightarrow u(j)$ of the variables. This implies it is invariant under the group $P(n)$ of all permutations of the variables. Thus, because of the linearity of the equation, one can classify solutions according to representations of $P(n)$. The two representations of interest in quantum mechanics are the one dimensional symmetric and antisymmetric representations. If a function belongs to the symmetric representation then it is invariant under the exchange $u(i) \leftrightarrow u(j)$ and if it belongs to the antisymmetric representation, it changes sign under the exchange $u(i) \leftrightarrow u(j)$. Bosons, which have integer spin, 0, 1, 2,…, , belong to the symmetric representation and fermions, which have odd half integer spin, 1/2, 3/2, …, belong to the antisymmetric representation.

## Gauge theory.

There is one more place where group representation theory occurs in quantum mechanics. In the quantum field theory formulation of quantum mechanics, the interactions are carried by gauge bosons. If we suppose the internal symmetry group is SU(n) (n≈6) then the generators of the group are the $n^2-1$ hermitian matrices $\tau^r$, r=1,…,$n^2$-1. The gauge fields associated with SU(n), $A_\mu^r(x)$, then transform under SU(n) just like the $\tau^r$ [9]. (The $\tau^r$ correspond to the adjoint

$n^2-1$ dimensional representation of SU(n).) The requirement that the theory is gauge-invariant then essentially completely determines the form of the allowed interactions. Thus in addition to specifying that the *static properties* of the constituents of matter are m,E,p,S,$s_z$,Q, group representation theory, through the gauge fields, plays a large role in determining the form of the *interactions*. The point is that **group representation theory permeates quantum mechanics at every turn.**

### Linearity, collapse and probability.

Finally, there is one place where linearity "fails." One can show that a theory in which the time evolution is both linear and unitary is not sufficient, by itself, to account for the probability law. Why? Because there can be no *probability of perception* in the mathematics [10]. No version of reality is singled out as perceptually special so the perceptions of every version of the observer are, in the mathematics, equally valid. Every version of reality is perceived on every run with 100% "probability" by one of the versions of the observer.

Because of this failure, one might advocate a collapse process governed by a linear but non-unitary time evolution. But we show in Appendix A that that cannot be; a mathematical collapse theory must be non-linear. So if one advocates a mathematical collapse mechanism, one must explain why, in spite of the non-linearity, all the results of group representation theory—which is entirely dependent on linearity—still hold.

One other remark. If one assumes there is collapse (even though there is no evidence for it), then the physical universe consists of the single, collapsed version of reality. That is, reality consists of wave functions/state vectors alone. In that case, although it is not normally explicitly stated, it is implicitly assumed that the wave functions carry the particle-like properties of mass, energy, momentum, spin and charge (because there is nothing else to carry them)!

## 6. No evidence for objectively existing particles.

One of the consequences of the arguments in Secs. 2-5 is that they imply there is no evidence for the objective existence of particles, separate from the wave function, where "objective" is taken here to mean there is only a single version of the particle, localized to a single, small region. This is shown by considering all the alleged evidence for particles and observing that all the particle-*like* characteristics of matter can be explained by the properties of the state vectors alone.

> 0. Before starting with the reputed evidence, we note that all the **numerical successes** of quantum mechanics, such as the prediction of the energy levels of hydrogen, come from equations for the wave functions (or state vectors; or fields). There are no equations for the time evolution of the positions and momenta of particles, for example. So in the mathematics at least, none of these successes give any hint that particles exist apart from the wave function.

> 1. Group representation theory shows that the particle-like properties of **mass, energy, momentum, spin, and charge** can be attributed to the state vectors. Thus it is not necessary to postulate the existence of particles as possessors or carriers of these attributes. (Sec. 5)

2. Even though there can be many versions of reality in a state vector, it is not necessary to assume the existence of single-version particles to explain why we **perceive only a single version** of reality. Linear quantum mechanics, by itself, implies the observer will never perceive more than one version. (Sec. 2)

3. Even though a wave function may be spread out over many grains of film, quantum mechanics predicts we will perceive only one localized grain as exposed. This explains the perceived **localization** of effects from spread-out wave functions. (Sec. 4)

4. An extension of the localization argument shows that quantum mechanics predicts we will perceive a sharply defined, unambiguous **particle-like trajectory** in a cloud chamber from a spread-out wave function. (Sec. 4)

5. The **photoelectric and Compton effects** follow from quantum mechanics alone because the state vectors corresponding to electrons and photons carry energy and momentum, and these quantities are conserved. (Secs. 2-5)

6. There are **thermodynamic and chemical** arguments which seem to imply the existence of particles. But they all involve the quantization (or unit-ization) of matter, and that occurs in pure quantum mechanics—a different wave function for each particle-*like* unit of matter—as well as in (reputed) particle-based physics.

7. Finally, if there are no particles, then one has the added bonus that there is no mystery in understanding the uncertainty principle (which is just a mathematical theorem concerning the properties of the wave function [11]), the non-local results of the Bell-Aspect experiment [12,13], Wheeler's delayed choice experiment [14], the quantum eraser [15] and other similar experiments. The properties of the state vectors alone correctly predict all the results of these experiments perfectly well, with no need to postulate action at a distance, the effect before the cause, and so forth.

These items constitute all the alleged evidence for particles. But since quantum mechanics alone can account for all these particle-*like* properties of matter, there is no reason to postulate the existence of objectively existing particles. It is still convenient to use the term "particle," but now instead of it referring to a single-version, localized, objective piece of matter, it refers to a multi-version state vector. The state vector corresponding to an electron will have mass $m_e$, charge $-e$, and spin ½ for example, while the state vector corresponding to a photon will have mass 0, charge 0, and spin 1.

The same argument would apply to "fields." There is no evidence for an objectively existing, single-version field—electron field, photon field—which has effects limited to a single small localized region.

We note that Secs. 2 through 6 essentially constitute a defense of the Everett many-worlds "interpretation" [16], with several arguments not found in Everett's paper filled in. However, the observation that linearity can explain so many seemingly puzzling effects does not mean that the linear laws of quantum mechanics alone constitute a valid interpretation. Why? Because as we remarked before, a linear, unitary theory alone cannot account for probability or the probability law [see Ref. 10 and Appendix A]. Each version is perceived on every run (in the mathematics) so there can be no *probability* of perception.

One more thought. If there are no objectively existing particles or fields, what does the ket |m,E,p,S,s_z,Q⟩ stand for? We will consider one possibility in the next section.

## 7. A potential form for an underlying theory

The Standard Model does an excellent job of describing the properties of elementary particles, especially since the mass-inducing Higgs boson has been detected [17, 18]. But at this point it could not be considered a theory which is satisfactory in all respects. It appears that some major modification is required. The usual approach is to look for a description of matter in terms of more and more elementary particulate constituents. But it may well be that that is not attacking the difficulties at the proper level; it is possible we need to look for a deeper, more fundamental form for quantum mechanics.

To obtain some idea of what such a theory might look like, suppose we consider only the *mathematical structure* of current quantum mechanics and for the time being pay no attention to the fact that it is meant to describe an actual physical reality. What is the most important mathematical characteristic of the current theory? I would say it is the pervasive group representational structure. Its basic constituents are the kets |m,E,p,S,s_z,Q⟩, which are basis vectors for representations of the direct product of ISL(2), the inhomogeneous Lorentz group, and $G_{int}$, the internal symmetry group. Group representations also enter because the fermion antisymmetric states and the boson symmetric states both correspond to one dimensional representations of the permutation group. And the gauge fields transform as the adjoint representation of the internal symmetry group. So there is hardly any basic part of the current quantum mechanical scheme that does not depend on group representation theory.

### The proposed basic structure of an underlying theory.

How might such a representational structure come about? One way is to suppose the current form of quantum mechanics is **the representational form of a pre-representational theory**. We will sketch such a possibility here. Assume there is a set of underlying variables $U_i$, $i=1,…,N$, such that one can define a group of transformations of those variables which is homomorphic to ISL(2)⊗$G_{int}$. Assume further there is a linear operator $O(U_i)$—a second order partial differential operator for example—which is invariant under this group and take

$$O(U_i)\Psi(U_i) = 0 \qquad (13)$$

as the linear equation at the heart of quantum mechanics, with the sets of variables $U_i$ and the linear operator $O$ currently unknown. Every physically allowed state—including the vacuum and states with any number of fermions and bosons—must correspond to a solution of this single equation. The equation is **pre-representational** in that $O$ does not involve the representational quantities m, E,p,S,s_z,Q; these are labels on the *solutions*. And it does not necessarily involve space-time variables because these will emerge from the group structure. The current form of quantum mechanics is to be a representation of this equation. Because of the invariance group, it is presumed that one can find basis vectors $\Psi_{m,E,p,S,sz,Q}(U_i) \equiv $ |m,E,p,S,s_z,Q⟩$_i$ which will be used to convert from the pre-representational form to the current representational form of quantum mechanics.

In addition to explaining the group representational structure of current quantum mechanics, there is an additional advantage to an underlying linear equation formulation; it can always be converted to a variational problem by observing that

$$\frac{\delta \langle \Psi | O | \Psi \rangle}{\delta \langle \Psi |} = 0 \Leftrightarrow O | \Psi \rangle = 0 \tag{14}$$

Presumably this variational formulation of the linear equation can be related to the Lagrangian variational principle of quantum field theory.

### A "single-particle" example of an underlying theory.

To make the ideas somewhat less abstract, it is helpful to give an example [2,3]. We will use a second-order differential equation in complex variables. Why complex? Because it is easier to include complex internal symmetry groups such as SU(n). These underlying (or "independent") variables $U_i$ do not correspond to anything currently known or conjectured in physics, including "hidden" variables. (The values of hidden variables determine the outcome of an experiment while the values of the $U$s do not.) The physically relevant states will be those functions of the underlying variables which are solutions to the underlying linear equation, and which have the appropriate group representation labels and properties.

The equation we choose as an example (because it is solvable) involves a harmonic-oscillator like operator;

$$O^{[1]} \Psi = 0 \tag{15}$$

$$O^{[1]} = -\left(\frac{\partial}{\partial u_{1i}} \frac{\partial}{\partial v_{2i}} - \frac{\partial}{\partial v_{1i}} \frac{\partial}{\partial u_{2i}} + \frac{\partial}{\partial \bar{u}_{1i}} \frac{\partial}{\partial \bar{v}_{2i}} - \frac{\partial}{\partial \bar{v}_{1i}} \frac{\partial}{\partial \bar{u}_{2i}}\right) \\ + (u_{1i} v_{2i} - v_{1i} u_{2i} + \bar{u}_{1i} \bar{v}_{2i} - \bar{v}_{1i} \bar{u}_{2i}) \tag{16}$$

There is a sum on $i$ from 1 to $n$, with $n$ on the order of 6, and the bar denotes complex conjugation.

### Symmetry Operations.
### The Lorentz Group

The underlying variables $U$ are $u_{1i}, v_{1i}, u_{2i}, v_{2i}$. There are 2 real variables for each $u_{1i}, v_{1i}, u_{2i}, v_{2i}$ for a total of 8 per $i$. Thus there are a total of $8n$ real variables. If we switch to variables in which $O$ is diagonal, it will be a quadratic form with half the coefficients +1 and half the coefficients –1. Thus the $O$ of Eq. (16) has a symmetry group with on the order of $32n^2$ generators. For now, we are only interested in those ten generators that correspond to the inhomogeneous Lorentz group. They must be hermitian $((\partial u)^* = -\partial \bar{u})$ and satisfy the commutation relations

$$[J_i, J_j] = i\varepsilon_{ijk} J_k, \quad [J_i, K_j] = i\varepsilon_{ijk} K_k, \quad [K_i, K_j] = -i\varepsilon_{ijk} J_k,$$
$$[J_i, P_j] = i\varepsilon_{ik} P_k, \quad [K_i, P_j] = i\delta_{ij} P_0, \quad [J_i, P_0] = 0, \quad [K_i, P_0] = iP_i \tag{17}$$
$$[P_i, P_j] = 0, \quad [P_i, P_0] = 0$$

The generators of the group *SL*(2), the set of all 2x2 complex matrices with determinant 1, homomorphic to the homogeneous Lorentz group, are

$$J_1 = \frac{1}{2}\left(u_{bi}\frac{\partial}{\partial v_{bi}} + v_{bi}\frac{\partial}{\partial u_{bi}} - \bar{u}_{bi}\frac{\partial}{\partial \bar{v}_{bi}} - \bar{v}_{bi}\frac{\partial}{\partial \bar{u}_{bi}}\right)$$

$$J_2 = \frac{i}{2}\left(-u_{bi}\frac{\partial}{\partial v_{bi}} + v_{bi}\frac{\partial}{\partial u_{bi}} - \bar{u}_{bi}\frac{\partial}{\partial \bar{v}_{bi}} + \bar{v}_{bi}\frac{\partial}{\partial \bar{u}_{bi}}\right)$$

$$J_3 = \frac{1}{2}\left(u_{bi}\frac{\partial}{\partial u_{bi}} - v_{bi}\frac{\partial}{\partial v_{bi}} - \bar{u}_{bi}\frac{\partial}{\partial \bar{u}_{bi}} + \bar{v}_{bi}\frac{\partial}{\partial \bar{v}_{bi}}\right) \quad (18)$$

$$K_1 = \frac{i}{2}\left(u_{bi}\frac{\partial}{\partial v_{bi}} + v_{bi}\frac{\partial}{\partial u_{bi}} + \bar{u}_{bi}\frac{\partial}{\partial \bar{v}_{bi}} + \bar{v}_{bi}\frac{\partial}{\partial \bar{u}_{bi}}\right)$$

$$K_2 = \frac{-1}{2}\left(-u_{bi}\frac{\partial}{\partial v_{bi}} + v_{bi}\frac{\partial}{\partial u_{bi}} + \bar{u}_{bi}\frac{\partial}{\partial \bar{v}_{bi}} - \bar{v}_{bi}\frac{\partial}{\partial \bar{u}_{bi}}\right)$$

$$K_3 = \frac{i}{2}\left(u_{bi}\frac{\partial}{\partial u_{bi}} - v_{bi}\frac{\partial}{\partial v_{bi}} + \bar{u}_{bi}\frac{\partial}{\partial \bar{u}_{bi}} - \bar{v}_{bi}\frac{\partial}{\partial \bar{v}_{bi}}\right)$$

where *b* is summed from 1 to 2 and *i* from 1 to *n*. And the generators of translations are

$$P_0 = u_{1i}\frac{\partial}{\partial \bar{v}_{2i}} - v_{1i}\frac{\partial}{\partial \bar{u}_{2i}} - \bar{u}_{1i}\frac{\partial}{\partial v_{2i}} + \bar{v}_{1i}\frac{\partial}{\partial u_{2i}}$$

$$P_1 = -u_{1i}\frac{\partial}{\partial \bar{u}_{2i}} + v_{1i}\frac{\partial}{\partial \bar{v}_{2i}} + \bar{u}_{1i}\frac{\partial}{\partial u_{2i}} - \bar{v}_{1i}\frac{\partial}{\partial v_{2i}} \quad (19)$$

$$P_2 = i\left(-u_{1i}\frac{\partial}{\partial \bar{u}_{2i}} + v_{1i}\frac{\partial}{\partial \bar{v}_{2i}} + \bar{u}_{1i}\frac{\partial}{\partial u_{2i}} - \bar{v}_{1i}\frac{\partial}{\partial v_{2i}}\right)$$

$$P_3 = u_{1i}\frac{\partial}{\partial \bar{v}_{2i}} + v_{1i}\frac{\partial}{\partial \bar{u}_{2i}} - \bar{u}_{1i}\frac{\partial}{\partial v_{2i}} - \bar{v}_{1i}\frac{\partial}{\partial u_{2i}}$$

Note that there is no scale in the momenta. One could multiply by any constant or even an SL(2) invariant function of the 1 variables and the commutation relations still work. The scale must come from the interaction (when it is put in).

We can also give the transformations macroscopically. The homogeneous Lorentz transformation *L(A)* corresponding to the 2x2 matrix *A* from *SL*(2) (the set of all 2x2 matrices with determinant 1) has the effect (we drop the subscripts here)

$$L(A)\begin{bmatrix}u\\v\end{bmatrix} = \begin{bmatrix}a_{11} & a_{12}\\a_{21} & a_{22}\end{bmatrix}\begin{bmatrix}u\\v\end{bmatrix} \quad (20)$$

while the effect of translations is

$$e^{iP_\mu x_\mu} u_1 = u_1$$
$$e^{iP_\mu x_\mu} v_1 = v_1$$
$$e^{iP_\mu x_\mu} u_2 = u_2 + ix_0\bar{v}_1 + ix_1\bar{u}_1 - x_2\bar{u}_1 - ix_3\bar{v}_1 \quad (21)$$
$$e^{iP_\mu x_\mu} v_2 = v_2 - ix_0\bar{u}_1 - ix_1\bar{v}_1 + x_2\bar{v}_1 - ix_3\bar{u}_1$$

## Space and time
Space and time can be defined as functions of the underlying variables. They must satisfy the equations

$$[P_0, x_\mu] = i\delta_{0\mu}, [P_j, x_0] = 0, [P_j, x_k] = -i\delta_{jk} \qquad (22)$$

A set of $x_\mu$ which satisfies these equations is

$$\begin{aligned}
x_0 &= \eta_{jk}(u_{j2}\bar{u}_{k1} + v_{j2}\bar{v}_{k1} + c.c.)/Z \\
x_3 &= \eta_{jk}(u_{j2}\bar{u}_{k1} - v_{j2}\bar{v}_{k1} + c.c.)/Z \\
x_1 &= \eta_{jk}(u_{j2}\bar{v}_{k1} + v_{j2}\bar{u}_{k1} + c.c.)/Z \\
x_2 &= i\eta_{jk}(-u_{j2}\bar{v}_{k1} + v_{j2}\bar{u}_{k1} - c.c.)/Z \\
Z &= [\bar{\eta}_{jk}(u_{1j}v_{1k} - v_{1j}u_{1k}) - c.c]/i \\
\eta_{kj} &= -\eta_{jk}
\end{aligned} \qquad (23)$$

where the $\eta_{jk}$ are constants. For the more general problem, where the character of the $U$s is still unknown, it should still be possible to define $x$s as functions of the $U$s. This divides the $U_i$ space, for each $i$, into constant-$x_\mu$ surfaces. That is, a space-time grid is superimposed on the $U$-space.

A solution to Eq. (13) will correspond to the physical state for all time. In the Schrodinger picture, one sets the function $\Psi$ on a constant time surface in $U$-space and then the Hamiltonian, deduced from the $O$ of Eq. (13), indicates how the solution progresses from one constant-t surface to the next.

## An *SU(n)* internal symmetry
To make the correspondence with current physics more complete, we introduce an *SU(n)* internal symmetry, where *SU(n)* consists of the set of all *n*x*n* unitary matrices with determinant 1. If the kets $|j\rangle$ transforms as the *n* representation of *SU(n)* and the kets $|\bar{j}\rangle$ as the $\bar{n}$ representation, then $|j\rangle|\bar{j}\rangle$, summed on *j* from 1 to *n*, is invariant under *SU(n)*. In order to make the generators of the inhomogeneous Lorentz group *SU(n)* invariants, we suppose the variables with $b=1$ transform as the *n* representation and the variables with $b=2$ transform as the $\bar{n}$ representation. The *SU(n)* transformation properties of all the variables and their derivatives are then

$$\begin{aligned}
|j\rangle &: u_{1j}, v_{1j}, \partial \bar{u}_{1j}, \partial \bar{u}_{1j}, \bar{u}_{2j}, \bar{v}_{2j}, \partial u_{2j}, \partial v_{2j} \\
|\bar{j}\rangle &: u_{2j}, v_{2j}, \partial \bar{u}_{2j}, \partial \bar{u}_{2j}, \bar{u}_{1j}, \bar{v}_{1j}, \partial u_{1j}, \partial v_{1j}
\end{aligned} \qquad (24)$$

So we see that the operator of Eq. (16) and the generators of the inhomogeneous Lorentz group are invariants under *SU(n)*. Charges will correspond to the elements of the $n-1$ diagonal generators of *SU(n)*. The $\eta_{jk}$ can be chosen so the $x_\mu$ are invariant under some subset of the SU(n) group.

## Particle-like States.

We will not go through the details of the mathematics, but one can show [3] that there are particle-like solutions to Eqs. (15,16). There are solutions for any nonzero mass and any half-integer spin. One of the unusual features is that the normalization involves a $\delta^4$ rather than the expected $\delta^3$;

$$\langle \psi_{p_\mu}(u,v) | \psi_{p'_\mu}(u,v) \rangle = \delta^4(p_\mu - p'_\mu) \tag{25}$$

The reason for this is that one of the generators of the invariance group obeys $[G, P_\mu] = i\lambda P_\mu$ so that it can be used to scale the mass. That is, if $\psi$ is a solution to the equation with mass $m$, then for any $\beta$, $e^{-i\beta G}\psi$ is also a solution (because $G$ commutes with $O$), and it has mass $\beta\lambda m$.

## A more general formulation.

We now suppose the set of all underlying variables consists of N "single-particle" sets $U_m$ and assume the linear operator has the form

$$\mathcal{O} = \sum_{m=1}^{N} O^{[1]}(U_m) + \sum_{m=1}^{N} \sum_{n \neq m}^{N} V(U_m, U_n) \tag{26}$$

with

$$V(U_m, U_n) = V(U_n, U_m) \tag{27}$$

The $O^{[1]}(U_m)$ are "single-particle" operators acting on functions of $U_m$, and the $V(U_m, U_n)$ are the interaction terms. Both $O^{[1]}$ and $V$ are assumed to be invariant under a group of transformations of the $U$s which is homomorphic to ISL(2)⊗$G_{int}$.

## Antisymmetry.

We see that $\mathcal{O}$ is invariant under exchanges $U_m \leftrightarrow U_n$ so it is invariant under the permutation group $P(N)$. Thus we can classify solutions according to the representations of $P(N)$. It is assumed that **all physical states** correspond to solutions belonging solely to the **totally antisymmetric** one-dimensional representation. So if $\Psi$ is such a solution, then

$$\Psi(U_1 \ldots U_m \ldots U_n \ldots U_N) = -\Psi(U_1 \ldots U_n \ldots U_m \ldots U_N) \tag{28}$$

## Single particle states. Kets.

Because of the invariance properties, we suppose there are solutions to $O^{[1]}\Psi = 0$ or perhaps $O^{[1]}\Psi = \lambda\Psi$ that can be labeled by the ISL(2) and $G_{int}$ representational labels $m, E, p, S, s_z, Q$. And then we can use the notation

$$\Psi_{m,E,p,S,s_z,Q}(U_i) \equiv |m, E, p, S, s_z, Q\rangle_i \tag{29}$$

That is, in this scheme a ket stands for a function of the underlying variables. The "elementary" fermion kets will presumably belong to the n or $\bar{n}$ representation of $SU(n)$ (if that is the actual gauge group).

The notation does not exactly match the usual ket designation because of the subscript $i$. But in the usual notation, it is tacitly understood that in the state, say, $|m_1, \ldots, Q_1\rangle|m_2, \ldots, Q_2\rangle$, the two kets correspond to two different "degrees of freedom." In the conventional scheme, the degrees of freedom would presumably correspond to the degrees of freedom of an actual particle. But we see from Sec. 6 that that correspondence cannot be supported. In the underlying variable scheme, the different degrees of freedom correspond to the different sets of underlying variables and are explicitly indicated.

We illustrate the construction of antisymmetric states in the underlying variable scheme by explicitly giving a simplified (no vacuum) two particle antisymmetrized state;

$$\Psi(m_1, \ldots, Q_1; m_2, \ldots, Q_2: U_1, U_2) =$$
$$\frac{[|m_1, \ldots, Q_1\rangle_1|m_2, \ldots, Q_2\rangle_2 - |m_1, \ldots, Q_1\rangle_2|m_2, \ldots, Q_2\rangle_1]}{\sqrt{2}} = \quad (30)$$
$$-\Psi(m_2, \ldots, Q_2; m_1, \ldots, Q_1: U_1, U_2) =$$
$$-\Psi(m_1, \ldots, Q_1; m_2, \ldots, Q_2: U_2, U_1)$$

We see that the minus sign occurs whether we exchange either underlying variable sets—that is, the subscripts on the ⟩ —or labels.

## Deriving quantum field theory from the underlying theory.

It is not our purpose here to present this model in detail; instead we will give a short summary of the possibilities. See [3] for more details.

**Vacuum state I.**
All states are to be antisymmetrized in the underlying variables so the vacuum state will be an antisymmetrized function of the $U_i$s. As a simple example, we might construct a Dirac vacuum out of products of functions of the negative energy "single particle" spin 1/2 functions of the $U_i$, with the total state antisymmetrized in the underlying variables.

**Fermion creation and annihilation operators.**
The appropriate *primary* basis vectors are taken to be functions of each set $U_i$ of underlying variables which transform as a spin 1/2 representation of ISL(2). To obtain a single particle state corresponding to a "physical" fermion, one multiplies the vacuum state, constructed from functions of N sets of underlying variables, by a positive energy single particle spin 1/2 function in the N+1st set of variables. Then one antisymmetrizes the state in all N+1 variables. A two-particle state will be constructed in essentially the same way, superimposing a second positive energy spin 1/2 state on the vacuum, with the state being antiysmmetrized in the N+2 variables. This process can be continued for any number of positive energy fermion states. One can then define fermion creation, *a\**, and annihilation, *a*, operators in the space of these

antisymmetrized states and, because of the antisymmetrization, they must obey the anticommutation relations

$$a^*(i)a(j) + a(i)a^*(j) = \delta_{ij} \tag{31}$$

In this way, we can define the antisymmetrized creation and annihilation operators of field theory for fermions. Fermion states then consist of fermion creation operators multiplying the vacuum. The use of creation and annihilation operators can be thought of as the way in which one deals with antisymmetric states in representation theory when the linear operator is invariant under *P*(N).

**Vacuum state II. Gauge transformations. Vector bosons.**

One could also consider a different kind of vacuum state. Change from the momentum specification of states to the $x_\mu$ specification of states by Fourier transforming. Then construct a "bound-state, localized molecule" of, say, 2n single-particle negative energy spin ½ states, with the molecule being invariant under *global* SU(n) transformations. (One would use non-invariant molecules for broken symmetry vacuum states.)

Because these bound state molecules are spread out in space-time, they will not be invariant under *local* (different on each $x_\mu$ surface) SU(n) transformations, which are to be identified with gauge transformations *G*(*x*). Instead, a gauge transformation will result in a disturbance of the vacuum molecules, with the disturbance being proportional to its space-time derivative, $\partial G/\partial x_\mu$ (assuming the molecule is very small compared to the variation of *G*). It is the disturbed molecules that correspond in this model to the interaction-carrying vector bosons. The disturbance is bilinear in the fermion creation operators making up the molecule, and because of this the creation and annihilation operators for the bosons will obey symmetric statistics. This agrees with gauge theory in which a gauge transformation produces a symmetrized boson field proportional to the space-time derivative of the gauge transformation.

One might be able to use gauge invariance as an aid to deducing the form of the underlying linear equation. Gauge invariance would presumably imply that if Ψ is a solution to Eq. (13), then Ψ transformed by a space-time dependent internal symmetry is also a solution.

In summary, this outlines a way in which the conjectured pre-representational underlying theory could lead to the current representational quantum field theoretic form of quantum mechanics.

## The meaning of the underlying theory.
## Probability.

What can it mean to have the basic constituents of matter be functions of some set of unknown variables which we cannot observe? For our purposes in this article, that question is beside the point because the logical scheme is as follows: When you look at the basics of linear quantum mechanics, you find very few concepts.

- There is a vacuum state.
- There are particle-like states labeled by group theoretic labels.
- These states are either antisymmetric or symmetric.

- There are linear equations of motion that these states (or fields) must satisfy. These are invariant under a suitable group of transformations; and they can be derived from a variational principle.

Aside from probability, nothing else is needed to explain the physical world. The spectrum of hydrogen, the hardness of diamond and the properties of elementary particles all follow from these few basic concepts. Perception of a single version of reality and the exposure of only one film grain by a spread-out wave function also follow. So if we can find a mathematical scheme that leads to these four concepts, then we have a viable underlying theory of quantum mechanics in the sense that it reproduces all observations except probability. That is, even though the objects of our perception, as well as our own bodies, "exist" in an abstract space, the abstractly constructed human brain will, for each branch of the wave function, still have its firing-neuron structure, and that is all that is needed to account for our perceptions.

> [Note: This existence in an abstract space is not really so different from the current situation where the abstract space in which quantum mechanics operates is constructed from the kets. It is just that we give a more specific form—functions of the $U$s—to the kets. Another way to say the same thing is that current quantum mechanics is not compatible with a non-abstract world view in which physical reality consists of actual objects moving in a three-dimensional space.]

The remaining problem with the underlying pre-representational approach—and it is a huge one—is how to deal with probability. There is no hint in the conjectured abstract, pre-representational form of the theory, just as there is none in the current representational form, that the perception of one version on each run is favored over perception of the other versions. And if there is no version whose perceptions are favored, it seems impossible to have *probability* of perception.

## 8. Conclusions.

There are two conclusions here, one on the non-existence of particles and the other on the possibility of a theory underlying quantum mechanics. First, the classical, Newtonian way of characterizing physical existence is that there is a single version of reality made up of localized particles. There is a good deal of reputed evidence for this view but they all boil down to just a few observations. We find all these observations can be explained using the principles—particularly linearity and group representation theory—of quantum mechanics.

**cl:** We perceive a single version of reality.
**qm:** Many versions of reality exist in the wave function. But the linearity of quantum mechanics implies the observer cannot perceive more than one version.

**cl:** After "an electron" scatters off "a proton," it is observed to travel in only one direction.
**qm:** The linearity of quantum mechanics implies that even though the electron wave function scatters in all directions, it will be perceived as traveling in only one direction.

**cl:** Particles carry or possess the properties of mass, energy, momentum, spin and charge.

**qm:** Group representation theory can be used to show that mass, energy, momentum, spin and charge are all properties of the wave function/state vector. It also shows that energy, momentum, spin and charge are conserved.

**cl:** A small part of a classical wave contains a correspondingly small part of the energy of the wave.
**qm:** In quantum mechanics, a small part of the wave carries the full energy.

The fact that quantum mechanics can be used to explain these basic particle-like properties implies there is no evidence for the objective existence of particles; all particle-like properties can be explained by properties of the state vectors alone.

      The second conclusion has to do with the extensive use of group representation theory in quantum mechanics. It is used to derive that mass, energy, momentum, spin and charge are attributes of matter. Symmetric and antisymmetric wave functions correspond to one-dimensional representations of the permutation group. Gauge fields transform as the adjoint representation of the internal symmetry group. And representation theory is an indispensable tool in the Standard Model. That is, every aspect of current quantum mechanics is constructed **exactly as if** it were a representation of an underlying pre-representational form of the theory.
      To illustrate what a pre-representational theory might look like, we suppose it consists of a single, linear partial differential equation in some currently unknown set $U_i$ of underlying variables. Neither space nor time nor the particle-like properties of mass, spin and charge enter the equation; these all correspond to properties of its solutions. All physical states—the vacuum, fermions, and bosons—are presumed to correspond to antisymmetrized solutions of this single equation, with the vacuum state composed of globally invariant "molecules" constructed from the spin ½ basis kets.
      It is shown in outline how quantum field theory, including a variational principle, can be derived by using group representation theory applied to the equation. It is presumed the linear operator $O(U)$ is invariant under a group of transformations of the underlying variables which is homomorphic to the direct product of the inhomogeneous Lorentz, a global internal symmetry group, and the permutation group defined by exchanges of variable sets $U_i$. This allows one to use spin 1/2 particle-like functions of the underlying variables as basis vectors. The presumed invariance allows us to superimpose a space-time grid on the space of the underlying variables. Gauge transformations are then space-time dependent internal symmetry transformations, and vector bosons—quadratic in the fermion states, so they obey symmetric statistics—are identified with gauge transformation-induced perturbations of the molecules making up the vacuum state.
      What are the potential advantages of an underlying theory? The one stressed here is that it explains the group representational structure of current quantum mechanics. But one would also hope that it would shed light on: why interactions take their vector boson-mediated form; why the vacuum has a particular broken symmetry and how that translates into the specific masses of the bosons and fermions; and what lies beyond the Standard Model, particularly for leptons. One might also hope that some insight would be gained into how gravity fits in. Perhaps it corresponds to a semi-macroscopic theory in which the local structure of the vacuum is slightly altered by the presence of a concentration of matter.

## Appendix A. Collapse, if it occurs, implies a non-linear theory.

Linear, unitary quantum mechanics cannot directly account for probability because there is no probability in the linear time evolution of the states. All possibilities "exist" in the wave function at every instant and no version of the observer is singled out by the mathematics as being more likely to correspond to our perceptions. To get around this problem, one might speculate (even though there is no evidence for collapse) that a linear, *non*-unitary theory could account for probability by causing collapse to just one version. To investigate this possibility, we consider mathematically based collapse schemes in which the collapse occurs smoothly as a function of time. We find that such a process cannot be linear.

The state at time 0, including the atomic states |k⟩ and detectors for each different possible outcome, is assumed to be a sum of n versions of reality

$$|\Psi(0)\rangle_m = \sum_{k=1}^{n} |k\rangle_m |D:k, yes\rangle_m \quad (A1)$$

The index *m* refers to the $m^{th}$ run of the experiment. There are n detectors, with

$$|D:k, yes\rangle_m \equiv |D1, no\rangle_m \ldots |Dk, yes\rangle_m \ldots |Dn, no\rangle_m \quad (A2)$$

Let T(t) be the linear but not necessarily unitary time translation operator. Because of the linearity, the state at time t can be written as

$$|\Psi(t)\rangle_m = T(t)|\Psi(t)\rangle_m = \sum_{k=1}^{n} a(k) T(t)[|k\rangle_m D:k, yes\rangle_m] \quad (A3)$$
$$= \sum_{k=1}^{n} a(k) \beta_m(k,t) |k,t\rangle_m |D:k, yes\rangle_m$$

where the $\beta_m(k, t)$ indicates that the normalization of state k may change because of the possible non-unitarity of T(t). (This pre-supposes there is some unknown mechanism whereby each version can evolve differently on different runs.) If there is collapse to one version of reality, then for each *m* one of the $\beta_m(k, t)$ must dominate as t becomes large. Thus the quantity

$$X_m(k,t) = \frac{|\beta_m(k,t)|^2}{\sum_{j=1}^{n} |\beta_m(j,t)|^2} \quad (A4)$$

must go either to 0 or to 1 on each run as *t* becomes sufficiently large. If the probability law is to be satisfied, it goes to 1 on a fraction $|a(k)|^2$ of the runs and 0 on all the other runs so that, for large N and t,

$$\frac{1}{N}\sum_{m=1}^{N} X_m(k,t) = |a(k)|^2 \tag{A5}$$

But because of linearity and the definition of the $\beta$s in Eq. (A3), the $\beta$s and therefore the $X$s do not depend on the coefficients $a(k)$. Thus Eq. (A5) cannot hold in a linear theory, so that any mathematically smooth collapse theory must be non-linear. (The basic problem is that collapse requires coordination between the changes in the coefficients and there can be no coordination between branches in a linear theory.)

So if, because of all its successes, we were to *require* that the theory of the physical universe be linear, then there can be no collapse.

We also note that there is no evidence that either collapse or hidden variables is the actual cause of probability.

## Appendix B. The Basics of Group Representation Theory.

We give here a brief review of the basic elements of representation theory.

### Invariance group of a linear operator.
As an illustration, consider the equation

$$\left(\frac{\partial}{\partial u}\frac{\partial}{\partial \bar{u}} + \frac{\partial}{\partial v}\frac{\partial}{\partial \bar{v}}\right)\psi(u,v) \equiv O\psi(u,v) = 0 \tag{B1}$$

where $u,v$ are complex variables, $O$ stands for the linear, partial differential operator, and the bar denotes complex conjugation. This equation is invariant under the set of all unitary transformations of the $u,v$ variables

$$\begin{bmatrix} u' \\ v' \end{bmatrix} = \begin{bmatrix} a_{11} & a_{12} \\ a_{21} & a_{22} \end{bmatrix} \begin{bmatrix} u \\ v \end{bmatrix}, \quad A^* = A^{-1} \tag{B2}$$

That is, one can show that

$$O(u',v') = \frac{\partial}{\partial u'}\frac{\partial}{\partial \bar{u}'} + \frac{\partial}{\partial v'}\frac{\partial}{\partial \bar{v}'} = \frac{\partial}{\partial u}\frac{\partial}{\partial \bar{u}} + \frac{\partial}{\partial v}\frac{\partial}{\partial \bar{v}} = O(u,v) \tag{B3}$$

If one unitary transformation is multiplied by another, the product is also a unitary transformation, so the set of all unitary 2x2 transformations forms the invariance group, SU(2), of the operator $O$.

### Basis vectors.
We define the operator $U(A)$ such that

$$U(A)f(u,v) = f(u',v') \tag{B4}$$

where $u', v'$ are defined in Eq. (B2). Then if $f$ is a solution to Eq. (B1), we have

$$U(A)[O(u,v)\psi(u,v)] = 0$$
$$= O(u',v')\psi(u',v') \qquad (B5)$$
$$= O(u,v)\psi(u',v')$$

That is, if the original function is a solution to the equation, so is the function with the variables transformed. But the set of all functions $\psi(u', v') = \psi(a_{11}u + a_{12}v, a_{21}u + a_{22}v)$, as A runs through SU(2), are not, in general, linearly independent. If we start with the solution $u^2$ for example, then as A runs through SU(2), there are only three independent functions— $u^2$, $uv$, and $v^2$. We can think of these three functions as forming the basis for a three dimensional vector space which is closed under the unitary transformations of Eq. (B2).

It is useful to introduce an SU(2)-invariant scalar product in this vector space. We choose

$$\langle f|g \rangle = \frac{2}{\pi^2} \int d[Re(u)] \int d[Im(u)] \int d[Re(v)] \int d[Im(v)] \bar{f}g \qquad (B6)$$
$$= \frac{2}{\pi^2} \int_0^1 |u|d|u| \int_0^{2\pi} d\theta \int_0^1 |v|d|v| \int_0^{2\pi} d\varphi \, \bar{f}g$$

We then represent the functional basis vectors by kets according to

$$u = |1/2\rangle, \quad v = |-1/2\rangle \qquad (B7)$$
$$\langle i|j \rangle = \delta_{ij}$$

That is, kets stand for functions of the underlying variables $u$ and $v$.

The three functions $u^2, uv, v^2$, or, in ket notation,

$$\sqrt{3/2}\, u^2 = |1\rangle, \quad \sqrt{2}uv = |0\rangle, \quad \sqrt{3/2}\, v^2 = |-1\rangle \qquad (B8)$$
$$\langle i|j \rangle = \delta_{ij}$$

are also carried into linear combinations of each other by the transformations of Eq. (B2). Remembering that the kets stand for function of $u,v$, we have, summing $j$ from 1 to 3, and using Eq. (B3),

$$U(A)|i\rangle = {}^{[3]}R_{ji}(A)|j\rangle \qquad (B9)$$

where the trailing superscript indicates the $R$s, quadratic in the $a_{ij}$, form a three dimensional **representation** of SU(2) in the sense that multiplication is preserved. That is, if transformation $A$ is followed by transformation $B$ and $BA=C$, then ${}^{[3]}R(B){}^{[3]}R(A) = {}^{[3]}R(C)$, where the multiplications are matrix multiplications. Similarly the four function $u^3, u^2v, uv^2, v^3$, suitably normalized, form the basis for a four dimensional representation of SU(2), and so on.

**Generators of transformations.**

The structure of continuous groups is almost completely determined by transformations very near the identity. To illustrate, we will use O(3), the group of all rotations in three dimensions. Consider small rotations, $\varepsilon$, about the z-axis,

$$x' = x\cos(\varepsilon) + y\sin(\varepsilon) \cong x + y\varepsilon$$
$$y' = y\cos(\varepsilon) - x\sin(\varepsilon) \cong y - x\varepsilon \qquad (B10)$$
$$z' = z$$

This transformation of variables has the following effect on an arbitrary function;

$$U_z(\varepsilon)f(x,y,z) \cong f(x + y\varepsilon, y - x\varepsilon, z)$$
$$\cong f(x,y,z) + i\varepsilon[ix\partial_y - iy\partial_x]f(x,y,z) \qquad (B11)$$
$$= f(x,y,z) + i\varepsilon L_z f(x,y,z)$$

where the linear, first order differential operator $L_z$ is the Hermitian **generator** of an infinitesimal rotation about the z axis. We can also construct generators for rotations around the x and y axes, with the results

$$L_x = i[y\partial_z - z\partial_y], \qquad L_y = i[z\partial_x - x\partial_z], \qquad L_z = i[x\partial_y - y\partial_x] \qquad (B12)$$

We find that the commutator of any two of these yields a generator back;

$$[L_x, L_y] = iL_z, \qquad [L_y, L_z] = iL_x, \qquad [L_z, L_x] = iL_y \qquad (B13)$$

More generally, every continuous group yields a finite number of linearly independent generators of infinitesimal transformations, and the commutator of any two of them is a linear combination of the generators. The commutation relations essentially define the group.

It is interesting that the three generators of SU(2),

$$S_x = (1/2)(u\partial_v + v\partial_u) + h.a.$$
$$S_y = (i/2)(v\partial_u - u\partial_v) + h.a. \qquad (B14)$$
$$S_z = (1/2)(u\partial_u - v\partial_v) + h.a.$$

obey the same commutation relations as the $L$s of Eq. ($B13$);

$$[S_x, S_y] = iS_z, \qquad [S_y, S_z] = iS_x, \qquad [S_z, S_x] = iS_y \qquad (B15)$$

Therefore the two groups, O(3) and SU(2), must be essentially identical (actually there is a 2 to 1 map of SU(2) onto O(3)). It is also of interest that the set of all 2x2 complex matrices of determinant 1, SL(2), is homomorphic to the homogeneous Lorentz group.

**Matrix representatives of linear operators.**
One can obtain a matrix representation of a linear operator, such as a generator, using the idea of Eq. ($B9$). To verify that it is a faithful representation, we must show that the proper multiplication rule holds. If $R_{ij}(A)$ represents the matrix elements of the representation of A, then we have, with a sum over repeated indices,

$$U(A)|i\rangle = |j\rangle R_{ji}(A) \tag{B16}$$
$$R_{ji}(A) = \langle j|U(A)|i\rangle \tag{B17}$$
$$U(B)U(A)|i\rangle = U(B)|j\rangle R_{ji}(A) = |k\rangle R_{kj}(B) R_{ji}(A) \tag{B18}$$

But $U(BA)|i\rangle = |k\rangle R_{ki}(BA)$ and so the proper matrix multiplication rule, $R(BA) = R(B)R(A)$, holds.

As an example, consider the linear operator $S_x$ of Eq. (B14) and use the two dimensional basis of Eq. (B7). We see that

$$S_x|1/2\rangle = (1/2)v = (1/2)|-1/2\rangle = R_{21}|-1/2\rangle \tag{B19}$$
$$S_x|-1/2\rangle = (1/2)u = (1/2)|1/2\rangle = R_{12}|1/2\rangle$$

and so the two dimensional matrix representative of $S_x$ is

$$S_x = \frac{1}{2}\begin{bmatrix} 0 & 1 \\ 1 & 0 \end{bmatrix} \tag{B20}$$

**Irreducible representations.**
The set of five functions, $u, v, u^2, uv, v^2$, are mapped into linear combinations of each other by a transformation from SU(2), so they form the basis for a representation of that group. But $u$ and $v$ alone also form the basis for a two dimensional representation, and $u^2, uv, v^2$ separately form the basis for a three dimensional representation. Thus the five dimensional representation is reducible. But the two dimensional and three dimensional representations cannot be further reduced. This idea of reducible and irreducible representations generalizes to representations of all groups.

**Invariants.**
The operator $L_x^2 + L_y^2 + L_z^2 = L^2$ from O(3) commutes with all three $L$s. This implies it is invariant under all rotations. All the basis vectors in a given irreducible representation are eigenvectors of $L^2$, with the same eigenvalue, so $L^2$ is a multiple of the identity in each irreducible representation. Basis vectors for different representations have in general different values for the eigenvalue.

This idea also generalizes. For each group, there will be polynomials in the generators which commute with all the generators and are therefore invariant under transformation from the group. Basis vectors for each irreducible representation will be eigenvectors of each invariant, and each different irreducible representation will have a different set of eigenvalues. Basis vectors within an irreducible representation are usually distinguished by their being eigenvectors of one or more of the generators. For example, basis vectors for SU(2) are usually taken to be eigenvectors of $s_z$. If the representation is of dimension n+1, the n+1 eigenvalues are $-n/2, -n/2+1, \ldots, +n/2$.

For the inhomogeneous Lorentz group, there are two invariants, one corresponding to mass and one to spin so representations are labeled by m and S. The vectors within a representation are labeled by energy, momentum, and z component of spin. The true internal symmetry group is not currently known, so we don't know the invariant operators. The charges, which are the eigenvalues of the diagonal generators, label basis vectors within a representation.